\documentclass{PoS}

\def\obj{QSO~2237$+$0305}

\title{Microlensing variability in the gravitationally lensed quasar     
\vspace*{1mm}  \obj\ $\equiv$ the Einstein Cross
 \thanks{Based on observations made 
 with the ESO-VLT Unit Telescope \#~2 Kueyen 
 (Cerro Paranal, Chile; Proposals 
 073.B-0243(A\&B),
 074.B-0270(A), 
 075.B-0350(A), 
 076.B-0197(A),
 177.B-0615(A\&B), PI: F. Courbin).}
 }

\ShortTitle{Microlensing variability in the Einstein Cross}

\author{\speaker{A. Eigenbrod},$^a$ F. Courbin,$^a$ D. Sluse,$^a$ G. Meylan$^a$,
and E. Agol$^b$ \\
\llap{$^a$}Laboratoire d'Astrophysique, Ecole Polytechnique F\'ed\'erale
de Lausanne (EPFL)\\ 
Observatoire de Sauverny, 1290 Versoix, Switzerland\\
\llap{$^b$}Astronomy Department, University of Washington\\ 
Box 351580, Seattle, WA 98195, USA\\
E-mail: 
\email{alexander.eigenbrod@epfl.ch},
\email{frederic.courbin@epfl.ch},
\email{dominique.sluse@epfl.ch},
\email{georges.meylan@epfl.ch},
\email{agol@astro.washington.edu}
}

\abstract{We present the results of the first long-term (2.2 years) spectroscopic monitoring
of a gravitationally lensed quasar, namely the Einstein Cross \obj. 

We spatially deconvolve deep VLT/FORS1 spectra 
to accurately separate the spectrum of the  lensing galaxy from the
spectra of the quasar images. Accurate cross-calibration of the 
observations at 31 epochs
from October 2004 to December 2006
is carried out using foreground stars observed simultaneously with the 
quasar.  The quasar spectra are further 
decomposed into a continuum component and several broad emission lines.

We find prominent microlensing events in the quasar images A and B,
while images C and D are almost quiescent on a timescale of a few months. 
The strongest variations are observed in the continuum, and their amplitude is larger in the
blue than in the red, consistent with microlensing of an accretion disk. Variations 
in the intensity and profile of the broad emission lines are also reported, 
most prominently
in the wings of the \ion{C}{III]} and in the center of the \ion{C}{IV} emission lines. 
During a strong microlensing episode observed in quasar image A, the broad 
component of the \ion{C}{III]} is more magnified than the narrow component. 
In addition, the emission lines with higher ionization potentials are 
more magnified than the lines with lower ionization potentials, consistent with 
the stratification of the  broad line region (BLR) infered from reverberation 
mapping observations. 
}

\FullConference{The Manchester Microlensing Conference: The 12th International Conference and ANGLES Microlensing Workshop\\
		 January 21-25 2008\\
		 Manchester, UK}

\begin{document}

\section{Introduction}

Most of the quasar microlensing studies so far are based exclusively on
broad-band photometric monitoring (e.g. Colley \& Schild 2003; Udalski et al.
2006). These observations are dominated by variations of the continuum, 
making it difficult to disentangle between variations of the continuum and
of the broad emission lines (BELs). 
Both regions are affected by microlensing, but
in different ways depending on their size.

Several theoretical studies show how
multiwavelength lightcurves can constrain the energy profile of the quasar
accretion disk and also the absolute sizes of the line-emitting regions (e.g.,
Schneider \& Wambsganss 1990; Agol \& Krolik 1999; Abajas et al. 1999;  Kochanek 2004). 
In order to investigate the
inner structure of quasars, we started the first long-term spectrophotometric
monitoring of a lensed quasar. Our target is  the Einstein Cross \obj, well known
for its microlensing induced variability. 
The spectral variations  of the  four   quasar images  are followed  
with  the Very Large   Telescope (VLT) 
from October 2004 to December 2006. 
In this contribution we summarize the results presented in
Eigenbrod et al. (2008). 
The full analysis of our monitoring data 
requires detailed  microlensing  simulations 
in order to constrain the quasar energy profile 
and  BLR size. These simulations will be the subject of future papers.


\section{Observations and data analysis}

We acquired our 31-epoch  observations   with   the FOcal Reducer   and low
dispersion Spectrograph (FORS1), mounted on 
the Unit Telescope \#~2  of the ESO Very Large
Telescope (VLT) located at Cerro Paranal (Chile).  We performed our observations
in the  multi-object spectroscopy (MOS) mode.   This
strategy allowed us to get simultaneous observations of the main target
and   of  four  stars    used  as   reference point-spread   functions
(PSFs). These stars  were used to  spatially deconvolve the  spectra
with the spectral version  of the MCS deconvolution algorithm (Magain et al.
1998, Courbin et al.   2000), as
well  as to perform accurate flux  calibration of the target spectra
from one epoch to another. 
Technical details are described in Eigenbrod et al. (2008). 

Different emission  features are known   to be produced  in regions of
different  characteristic  sizes. 
Emission features from smaller regions of the source are more highly variable due to 
microlensing than features emitted in more extended regions 
(e.g. Wambsganss et al. 1990).  
In order to study the variation
of each spectral  feature independently,   
we decompose the spectra into 
%
%
%
(1) a power law
continuum $f_{\nu}     \propto \nu^{\alpha_\nu}$, (2) a pseudo-continuum due to the merging of \ion{Fe}{II}
and \ion{Fe}{III} emission blends, and (3) an emission spectrum due to the
individual BELs.  

\section{Results}

\begin{figure}
\includegraphics[width=9.55cm, height=6.2cm]{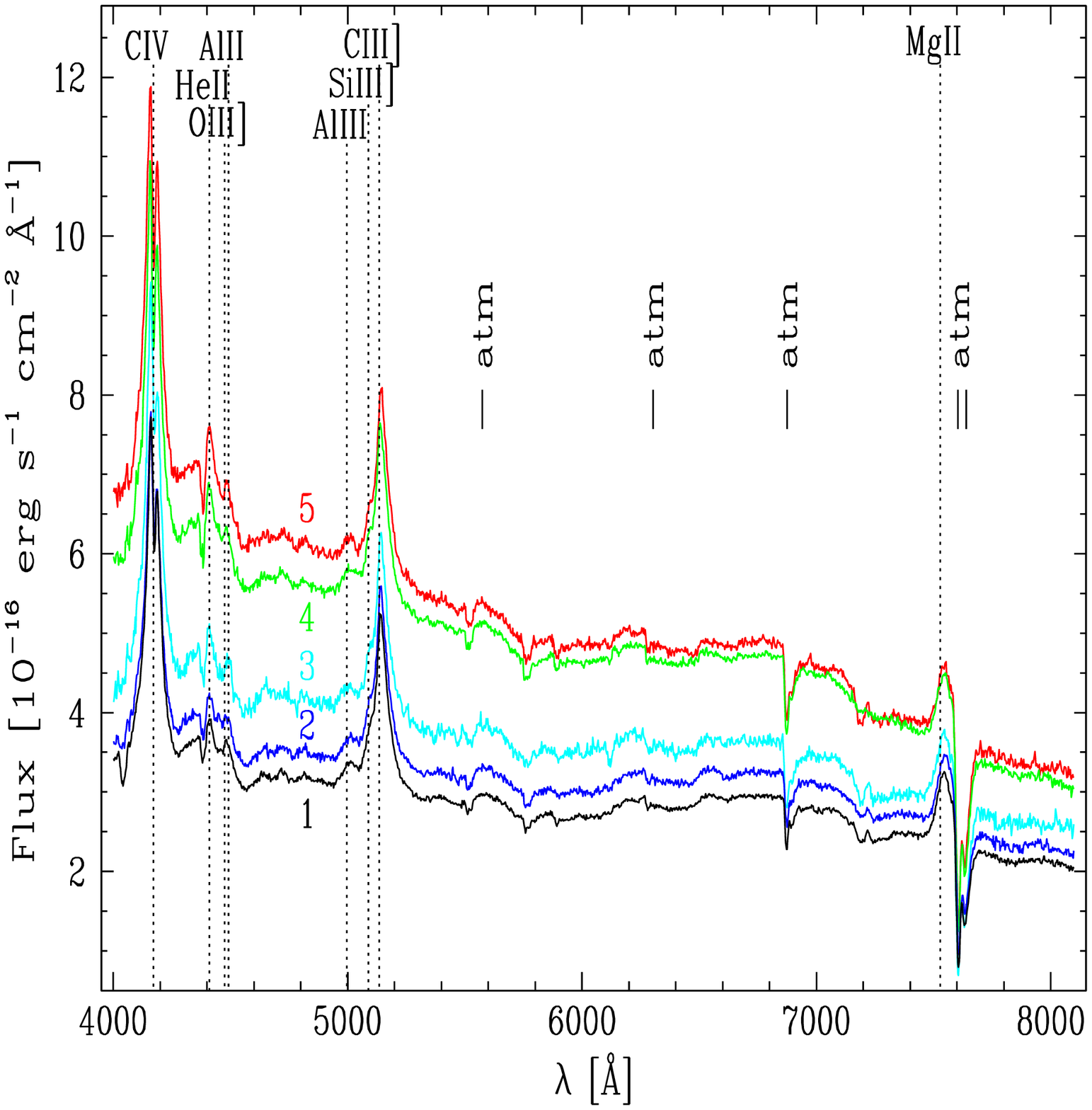}
\hspace{-0.7cm}
\includegraphics[height=6.2cm]{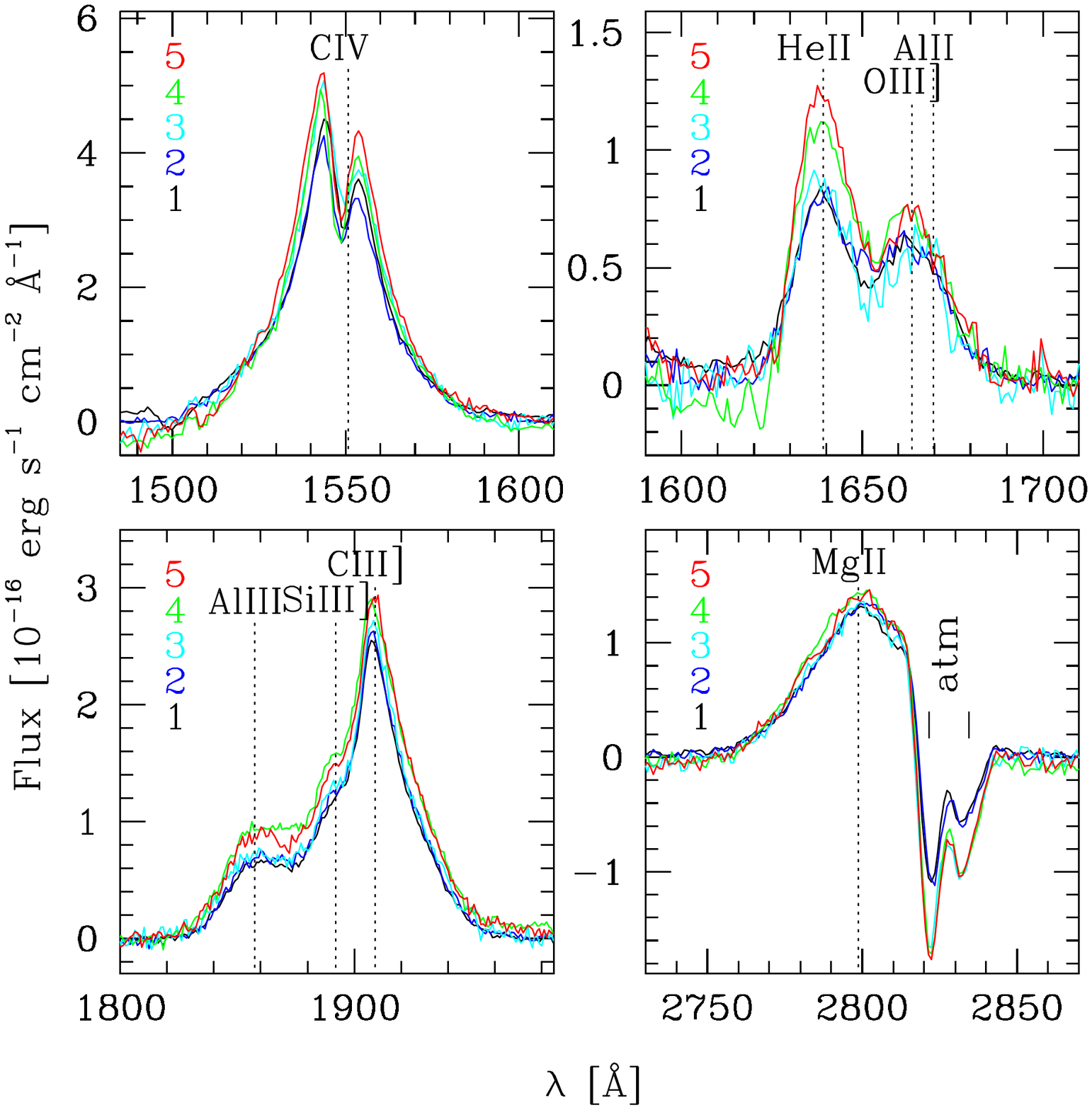}
\caption{Spectra of image A at 5 different epochs during the 
magnification event that occured during the summer 2006.
The epochs are chronological with
\#$1=$25-08-2005, 
\#$2=$21-10-2005, 
\#$3=$06-12-2005, 
\#$4=$27-07-2006, and 
\#$5=$26-11-2006. 
\emph{Left panel:} Chromatic variations of the continuum are conspicuous. The blue part
of the spectra are more magnified than the red part. This is consistent with
microlensing of an accretion disk.
\emph{Right panel:} We observe variations in the intensity and profile of the BELs, especially
in the wings of the \ion{C}{III]}, center of the \ion{C}{IV}, and intensity of the \ion{He}{II}
emission lines. In this last panel the continuum has been subtracted.}
\label{fig1}
\end{figure}

We observe chromatic  variations induced by microlensing in the
continuum of  all images of  \obj.  The effect is most
pronounced  in image  A during  the last  observing season (HJD$\sim$2453900~days), and in
image  B at  the   beginning of our  monitoring (HJD$\sim$2453500~days).
During these magnification events we notice a steepening of the 
continuum with the blue part of the spectra being
more magnified than the red part (see Fig.~\ref{fig1}). 
In quasar image A,
all the emission lines are less magnified than the continuum, consistent with a
scheme where the BLR is small enough to be significantly microlensed, but less
than the continuum, which is emitted  in a more compact region.
In addition, we find evidence the BLR is stratified, as lower ionization lines 
are less magnified and hence emitted in larger regions.
The same global trend is observed in image B.

The behaviour of the \ion{Fe}{II+III} emission is
more difficult to interpret, as it is a blend of many
lines. 
However, we note that 
the  difference  in  magnification of  the \ion{Fe}{II+III}
lines between a microlensing and a quiet  phase is larger than for the
other   lines.  This  may suggest  differential   magnification of the
emitting regions within the \ion{Fe}{II+III} complex, i.e. that the
\ion{Fe}{II+III} is present both in compact and in more extended regions,
a conclusion also reached by Sluse et al. (2007) with single-epoch spectra
of the quadruply imaged quasar RXJ~1131$-$1231.

We also  investigate  the
possibility of a change in the profile of the BELs.  
Such profile variations may
be  caused by  differential  magnification of  regions with  different
velocities in the BLR  
(e.g. Lewis 1998).  
%
In image A, 
the broadest part of the \ion{C}{III]} is
more microlensed than the core, indicating that the broadest component
of the \ion{C}{III]} emission is emitted in a more compact region than
the core of the emission line.
Moreover, we observe
nearly the same  magnification ratio between  the core and the broader
component of the \ion{C}{III]} emission line  than is reported between the narrow
\ion{[O}{III]} and  the broad \ion{H}{$\beta$} emission lines (Metcalf
et al.~2004). This gives a hint that 
the narrow component of the \ion{C}{III]} emission line 
is, at least in part, physically connected with 
the narrow line region (NLR).
In addition, we note evidence for variations in the central parts of
the \ion{C}{IV} emission   lines in  all  images and at
most  epochs.  

\section{Conclusions}

We present the first long-term spectrophotometric monitoring of a gravitationally lensed quasar; 
the Einstein Cross \obj. 
The mean temporal sampling is of one observation every second week.
The observations are carried out with
the VLT in a novel way, using the spectra of PSF stars, both 
to deblend the quasar images from the lensing galaxy and to carry out
a very accurate flux calibration.  


We find that all images of \obj\  are affected by microlensing. 
Image~A  shows an  important
brightening episode  at  the end of our monitoring campaign, and   image~B  at
the beginning. The continuum of these two images
becomes bluer as they get brighter, as expected from microlensing
magnification of an accretion disk.
We also report microlensing-induced variations of the BELs, both in
their integrated line intensities and in their profiles.  
In   quasar   image   A, the  broad   component   of  the
\ion{C}{III]} line is more magnified than the narrow
component. This might indicate that the core of this line
is emitted in the NLR.
%
%
Intensity variations in the BELs are detected mainly  in images A and
B. Our  measurements  suggest    that  higher ionization   BELs   like
\ion{C}{IV},  \ion{C}{III]},  are  more  magnified than lower   
ionization   lines like \ion{Mg}{II}. This  is  compatible with
reverberation mapping studies and  a stratified structure of  the BLR.
There    is  marginal evidence that   regions   of different sizes are
responsible for the \ion{Fe}{II+III} emission.

%
The very different behaviours of the BELs and the continuum with  res\-pect to
microlensing offer considerable hope to  reconstruct the  two types of
regions independently, using inverse ray-shooting simulations.


\section{Acknowledgements}
We are  extremely grateful to all ESO staff for their excellent work in
the acquisition of the observational data. 
This project is  partially supported  by the Swiss
National Science Foundation (SNSF).

\end{document}